\documentstyle[12pt,worldsci]{article}

\begin{document}
\bibliographystyle{unsrt}

\title{OFF-SHELL ELECTROMAGNETIC FORM FACTORS OF THE NUCLEON
IN CHIRAL PERTURBATION THEORY}

\author{J.W. BOS\\
{\em Physics Department, National Central University\\
Chung Li, Taiwan 32054}\\
\vspace{0.3cm}
and \\
\vspace{0.3cm}
J.H. KOCH\\
{\em NIKHEF-K, P.O. Box 41882\\
1009 DB Amsterdam, The Netherlands}}
\maketitle

\begin{abstract}
\footnotesize
We study the electromagnetic form factors of a  nucleon
in next-to-leading order chiral perturbation theory (CPT)
in the case where one of the
nucleons is off its mass shell.
We calculate the leading nonanalytic contributions
to relevant measures for the off-shell dependence
in the limited kinematical range allowed.

\end{abstract}

\section{Introduction}

Many investigations have been carried out recently to examine the
electromagnetic
interaction of a nucleon not on its mass shell. This was mainly done by
using nuclear reactions
such as inclusive electron scattering, (e,e$'$), and
nucleon knock-out, (e,e$'$p).
However, two-step processes on a free nucleon, such as Compton
scattering, allow one to probe several features of this interaction
under much simpler circumstances.
Since the electromagnetic vertex in these reactions involves a
nucleon which does not satisfy the free Dirac equation,
it is not correct to use the {\em free} nucleon electromagnetic
current operator.
The operator structure of
the off-shell $\gamma$NN vertex is more
complicated: Its most general form
consists of twelve Dirac operators, each multiplied by a
form factor which depends on three scalar variables\cite{binc60}.
As these off-shell form factors concern intermediate states,
they cannot directly be obtained from experimental data.
There have been efforts to obtain them under certain
assumptions from dispersion relations\cite{binc60,nyma70,hare72}.
However, because of bad convergence and the
absence of the necessary experimental
input, this approach was of little use in practice\cite{bos93.2}.
A microscopic model \cite{naus87,tiem90} for the internal
nucleon structure is needed to describe off-shell form factors,
and their explicit form depends on the specific representation
used\cite{sche94}.
Here we examine the general electromagnetic vertex of the
nucleon using the standard CPT lagrangian\cite{gass88,krau90}.

\section{The half off-shell  $\gamma$NN vertex}

The general form of the $\gamma$NN vertex
for the case that both
initial and final nucleon are off their mass shell,
has been given by Bincer\cite{binc60}.
We consider here the reducible
``half off-shell'' vertex with the initial nucleon
off-shell ($p^2=W^2$) and the final nucleon on-shell ($p'^2=M^2$),
which has the general decomposition
\begin{eqnarray}
	\overline{\rm u}(p') \Gamma^{\rm{N}}_{\mu}(p',p) & = &
	- {\rm i} e \overline{ \rm u}(p') \Bigl[
	( f_{1}^{++}
	\gamma_{\mu} + f_{2}^{++} \frac{{\rm i} \sigma_{\mu \nu}
	q^{\nu}}{2M}
	+ f_{3}^{++} q_{\mu} )
	\Lambda^{+}(p)
\nonumber
\\
	& & \mbox{} +
	( f_{1}^{+-}
	\gamma_{\mu} + f_{2}^{+-} \frac{{\rm i} \sigma_{\mu \nu}
	q^{\nu}}{2M}
	+ f_{3}^{+-} q_{\mu} )
	\Lambda^{-}(p) \Bigr],
\label{4:hovert}
\end{eqnarray}
where $f_{i}^{+\pm} = f_{i}^{+\pm} (q^2,W,M)$ and $\Lambda^{\pm}(p)$
are the projection operators
\begin{equation}
	\Lambda^{\pm} = \frac{\pm p\!\!\slash + W}{2W} .
\end{equation}
We will consider here only the form factor $f_1^{++}$.
Expanding around the photon point, one has
\begin{equation}
	f_1^{++}(q^2,W,M) = f_1^{++}(0,W,M) +
	q^2 \frac{\partial}{\partial q^2}f_1^{++}(q^2,W,M)|_{q^2
	= 0} + \ldots\, .
\label{4:paard}
\end{equation}
The first term in the expansion Eq.~(\ref{4:paard})
is defined by the nucleon charge, independent of $W$\cite{binc60}.
The $W$ dependence of $f_1^{++}$ will first show up in  the second
term, and we will therefore below consider
\begin{equation}
	6 \frac{\partial^2}{\partial W\,\partial q^2}
	f_1^{++} (q^2,W,M) |_{W=M,\,\,q^2=0} .
\label{4:q1}
\end{equation}
We have included the factor 6
since the second term in the right hand side of
Eq.~(\ref{4:paard}) is related
to the mean square radius of the nucleon and
the quantity in Eq.~(\ref{4:q1})
can be interpreted as the change of the mean square radius
with $W$.

\section{The $\gamma$NN vertex in CPT}

We calculate the electromagnetic interaction
term in next-to-leading
order CPT, i.e.\ up to order ${\cal O}( {E^3})$, where $E$ is a
generic notation for the four-momentum of the external photon, the
three-momenta of the external nucleons, and the meson mass\cite{gass88}.
For this purpose we need all tree-level diagrams from the
CPT lagrangian up to order ${\cal O}({E^3} )$, and
all one-loop diagrams from the
CPT lagrangian up to order ${\cal O}({E} )$, contributing to the vertex.
For off-shell particles one needs to take into account
more terms in the lagrangian than in the standard lagrangian.
In the usual notation\cite{gass88},
the additional terms to be added to the lagrangian are\cite{bos93}
\begin{eqnarray}
	\Delta{\cal L}_{\rm CPT} & = &
	n_1 \frac{M}{F^2} \bar{\psi}
	({\rm i} D\!\!\!\!\slash - M)^2 \psi
	+ n_2 \frac{1}{F^2} \bar{\psi} \{ ({\rm i} D\!\!\!\!\slash - M),
	\sigma^{\mu \nu} f_{\mu\nu}^+ \} \psi
\nonumber\\
	& & \mbox{}
	 + n_3 \frac{1}{F^2} \bar{\psi}
	({\rm i} D\!\!\!\!\slash - M)^3 \psi.
\label{4:mus}
\end{eqnarray}
The constants $n_1$ and $n_3$ in
Eq.~(\ref{4:mus}) are needed to renormalize the loop
diagrams for the off-shell
{\em irreducible\/} $\gamma$NN vertex and the nucleon
self-energy, while $n_2$ is finite.
These constants can contribute to a {\em physical\/} amplitude
where this half off-shell vertex plays a role, e.g.\
Compton scattering on a free nucleon.
Here, we will focus
on the leading singular terms in the chiral limit,
$M_{\pi} \rightarrow 0$, contributing
to Eq.~(\ref{4:q1}).
It has been argued\cite{gass88}
that all such nonanalytic terms are given
by a one-loop calculation.
Since the actual pion mass is small,
these contributions can be expected to be the dominant.
Analytic contributions to Eq.~(\ref{4:q1}) are outside the scope
of this calculation\cite{bos93}.

\section{Results and discussion}

For the {\em isovector\/} form factor $f_1^{++, {\rm V}}$
we find at $W=M$ and $q^2 = 0$
\begin{equation}
	6 \frac{\partial^2}{\partial
	W \partial q^2} f_1^{++, {\rm V}}|_{W=M,\,\,q^2=0}  =
	\frac{5}{32 \pi}
	\frac{g_{\rm \scriptscriptstyle A}^2}{F^2 M} \frac{1}{\sqrt{x}}
	+ {\cal O}( {\ln x}) .
\label{4:geit1}
\end{equation}
In the chiral limit this quantity diverges.
{}From Eq.~(\ref{4:geit1}) we can see
that the isovector mean square radius of a nucleon
with $W<M$ is smaller
than that of a free one.
This is in qualitative agreement with earlier results
\cite{naus87,tiem90}.
Taking $x$ at its physical value, $x = 0.022$, and retaining only
the leading terms in Eq.~(\ref{4:geit1}), we find in CPT that the free
($W = 940 \; {\rm MeV}$)
mean square radius of $0.6 \; {\rm fm}^2$ is reduced
by about $0.1 \; {\rm fm}^2$ if
we use Eq.~(\ref{4:geit1}) to extrapolate
linearly to $W = 890 \; {\rm MeV}$.
The singular term for the {\em isoscalar\/} form factor
$f_1^{++, {\rm S}}$ vanishes:
\begin{equation}
   6 \frac{\partial^2}{\partial
   W \partial q^2} f_1^{++, {\rm S}}|_{W=M,\,\,q^2=0} = {\cal
O}({1} ) .
\end{equation}
Its expansion starts with a ${\cal O}( {1})$ term, indicating a weaker
off-shell dependence than the isovector part.
Analytic contributions (which are now
the leading one) cannot
be obtained from our calculation.

The exact value of the quantities considered
depends on the size of the higher-order contributions. However,
our calculation is able to show  qualitative
features of the electromagnetic properties of a nucleon not on
its mass shell. Our calculation shows that these properties,
such as the ``radius of an off-shell nucleon,'' are not fixed
properties, but depend on the kinematical parameters and dynamical
circumstances.
Results for the other form factors in Eq.~(\ref{4:hovert}) and
a more detailed discussion of the calculation can be found
elsewhere\cite{bos93}.

\end{document}